\documentclass[aps,prb,twocolumn,groupedaddress,showpacs]{revtex4}
\usepackage{graphicx}
\usepackage{color}
\usepackage{epstopdf}
\usepackage{amsmath}
\usepackage{multirow}
\usepackage{dcolumn}
\usepackage{bm}

\begin{document}

\title{Finite-size correction in many-body electronic structure calculations\\ of magnetic systems}
\author{Fengjie Ma, Shiwei Zhang, and Henry Krakauer}
\affiliation{Department of Physics, College of William and Mary, Williamsburg, VA 23187.}

\begin{abstract}

We extend the post-processing finite-size (FS) correction method, developed by Kwee, Zhang, and Krakauer [{\em Phys. Rev. Lett.} {\bf 100}, 126404 (2008)], to spin polarized systems. The method estimates the FS effects in many-body electronic structure calculations of extended systems by a modified density functional theory (DFT) calculation, without having to repeat expensive many-body simulations. We construct a unified FS DFT exchange-correlation functional for spin unpolarized and fully spin polarized systems, under the local density approximation. The results are then interpolated to arbitrary spin polarizations. Generalization to other functional forms in DFT are discussed. The application of this FS correction method to several typical magnetic systems with varying supercell sizes demonstrates that it consistently removes most of the FS errors, leading to rapid convergence of the many-body results to the infinite size limit.

\end{abstract}


\maketitle

\section{Introduction}

Many-body simulation methods, such as diffusion Monte Carlo (DMC) \cite{ca,reynolds,qmc} and auxiliary field Quantum Monte Carlo (AFQMC), \cite{afqmc} are capable of yielding highly accurate results for electronic systems. In extended systems, however, because of the use of periodic boundary conditions (PBC), finite-size (FS) errors arise in many-body (MB) calculations, which are often larger than the statistical and other systematic errors. The 
FS error reflects the spurious interactions of the system, modeled by a finite-size simulation cell (supercell), with its own periodic images. The long-range nature of the Coulomb interaction
in electronic systems exacerbates the problem, and makes the FS error more pronounced.  
To reduce this error, calculations have to be performed using larger and larger simulation cells, in order to extrapolate the results to the infinite limit of the supercell size. 
The computational cost of MB calculations is typically high.
For example, QMC methods
scale with system size as $N^3$-$N^4$, with large prefactors. 
Other MB methods typically have significantly worse scaling.
Thus convergence of MB calculations with supercell size is difficult to achieve, and 
accurate extrapolation is often too costly. \cite{williamson1}
FS correction schemes that can accelerate this convergence are therefore highly desirable.

The behavior of the FS error in a MB calculation is different from that in a standard density-functional theory (DFT)
calculation \cite{hk,ks} (or other independent-electron calculations), as we further discuss in 
Sec.~\ref{sec:prel}. The MB FS error consists of a part that
is essentially parallel with the corresponding FS error in a DFT calculation of the same 
supercell. This part, which will be termed the one-body (1B) FS error, is easier to correct.
The residual error, which will be termed the two-body (2B) FS error,
 is due to the approximate treatment of electron interactions in the extended
system by the electrons in the simulation cell plus their images. 
The 1B and 2B errors are largely separable (see Sec.~\ref{sec:prel}). The correction of 
the 2B errors is the focus of FS correction methods.

It has been shown \cite{fraser,kent,williamson,drummond} that using a modified periodic Coulomb potential instead of the standard 
Ewald form of the electron-electron repulsion can significantly reduce FS errors in the 
calculated ground-state energy and accelerate 
convergence. More recently, Chiesa {\em et al.\/} \cite{chiesa} introduced a 2B FS correction based on the random phase approximation analysis of the momentum distribution and 2B structure factor, which achieves similar results. Both of these approaches are internal correction methods. The former requires 
modification of the form of the electron-electron Coulomb interaction, and the 
latter requires calculation of the structure factor in the MB calculation.

The method of Kwee, Zhang, and Krakauer (KZK) is an external FS correction approach. \cite{kzk} The idea is to
have independent-electron calculations which mimic the FS behavior of MB calculations. KZK developed a FS exchange-correlation (XC) functional which describes the effect of electron-electron interaction in the spirit of a local-density approximation (LDA) in DFT
and which takes into account the finite size of the simulation cell.  A 
functional was parametrized for spin-unpolarized systems. The functional has been included in the software package 
Quantum Espresso \cite{pwscg} and the method is applicable to any MB total energy results \cite{kzk,wirawan,sola,maezono,binnie,hennig,sorella} as a simple, post-processing correction.

The FS functional of KZK does not include spin-dependence, however. In this paper we generalize
the method to arbitrarily spin-polarized systems, and parametrize a FS XC functional 
based on the local spin-density approximation (LSDA). The new functional includes KZK, but 
will allow FS corrections in 
systems with spin-polarization or magnetic order, for example in many transition-metal oxide 
solids. In such systems the effect of electron correlation is generally stronger, and the need for 
MB calculations is greater. The complexity of the systems also means our ability to systematically 
simulate larger and larger supercells will be more limited. 
Accurate and simple FS corrections will therefore be very valuable. In the applications we 
illustrate how the new FS LSDA functional can help accelerate convergence of the MB 
results in various systems.

The remainder of the paper is organized as follows. In Sec.~\ref{sec:prel}, we describe 
the overall formalism of our FS correction and define some notations. In Sec.~\ref{sec:XC},
we present the FS XC
functionals, with the exchange and correlation parts each in a separate 
subsection, and a summary of the parameter values of the result of our parametrization in the 
last subsection. In Sec.~\ref{sec:apps}, several applications are presented to illustrate the 
FS correction method. We further discuss several aspects of the method in Sec.~\ref{sec:discuss}
before summarizing in Sec.~\ref{sec:summary}.

\section{Preliminaries}
\label{sec:prel}

We will consider zero-temperature total energy calculations in a cubic supercell of side length 
$L$, with $M$ atoms in it. 
Twist boundary conditions \cite{lin,kzk} will usually be applied, 
the many-body generalization of ${\mathbf k}$-points in independent-electron calculations.
The FS error for a given twist ${\mathbf k}$ is defined as 
\begin{equation}
\Delta E({\mathbf k},L)\equiv E^\infty-E({\mathbf k},L),
\label{eq:def_FSerror}
\end{equation}
where each $E$ denotes \emph{per atom} energy, and $E^\infty$ is the corresponding result
at the infinite supercell limit.

The FS error can be separated into 1B and 2B parts, which are also referred to as independent-particle and Coulomb FS errors. \cite{kent} The 1B FS error arises 
from incomplete ${\mathbf k}$-point integration, i.e., a discrete momentum space mesh:
\begin{equation}
\Delta E^{\rm 1B}({\mathbf k},L)\equiv \sum_{\{{\mathbf k}\}}E({\mathbf k},L) - E({\mathbf k},L),
\label{eq:def_1B}
\end{equation} 
where in the first term on the right 
the sum is over a sufficiently large set of ${\mathbf k}$-points 
(with appropriate weight if necessary)
to reach convergence. Below we will also use the notation $\Delta E(\{{\mathbf k}\},L)$ to denote this term.
In a standard DFT calculation, because of  Bloch's theorem,
the 1B FS error is the only form of FS error:
\begin{equation}
E_{\rm DFT}^\infty=E_{\rm DFT}(\{{\mathbf k}\}\rightarrow\infty,L)
=E_{\rm DFT}(\{{\mathbf k'}\}\rightarrow\infty,L'),
\label{eq:DFT_FSerror}
\end{equation}
{\it i.e.}, $\Delta E_{\rm DFT}({\mathbf k},L)=\Delta E^{\rm 1B}({\mathbf k},L)$  in DFT.
In Eq.~(\ref{eq:DFT_FSerror}), different choices of the simulation cell lead to different requirements
on the set $\{ {\mathbf k}\}$ over which to sum, 
but for each choice of $L$, the infinite limit 
can be achieved with enough ${\mathbf k}$-points. The usual approach is to 
simply integrate over a dense ${\mathbf k}$-point grid
using the primitive cell. 

\begin{figure}
\includegraphics[width=0.43\textwidth]{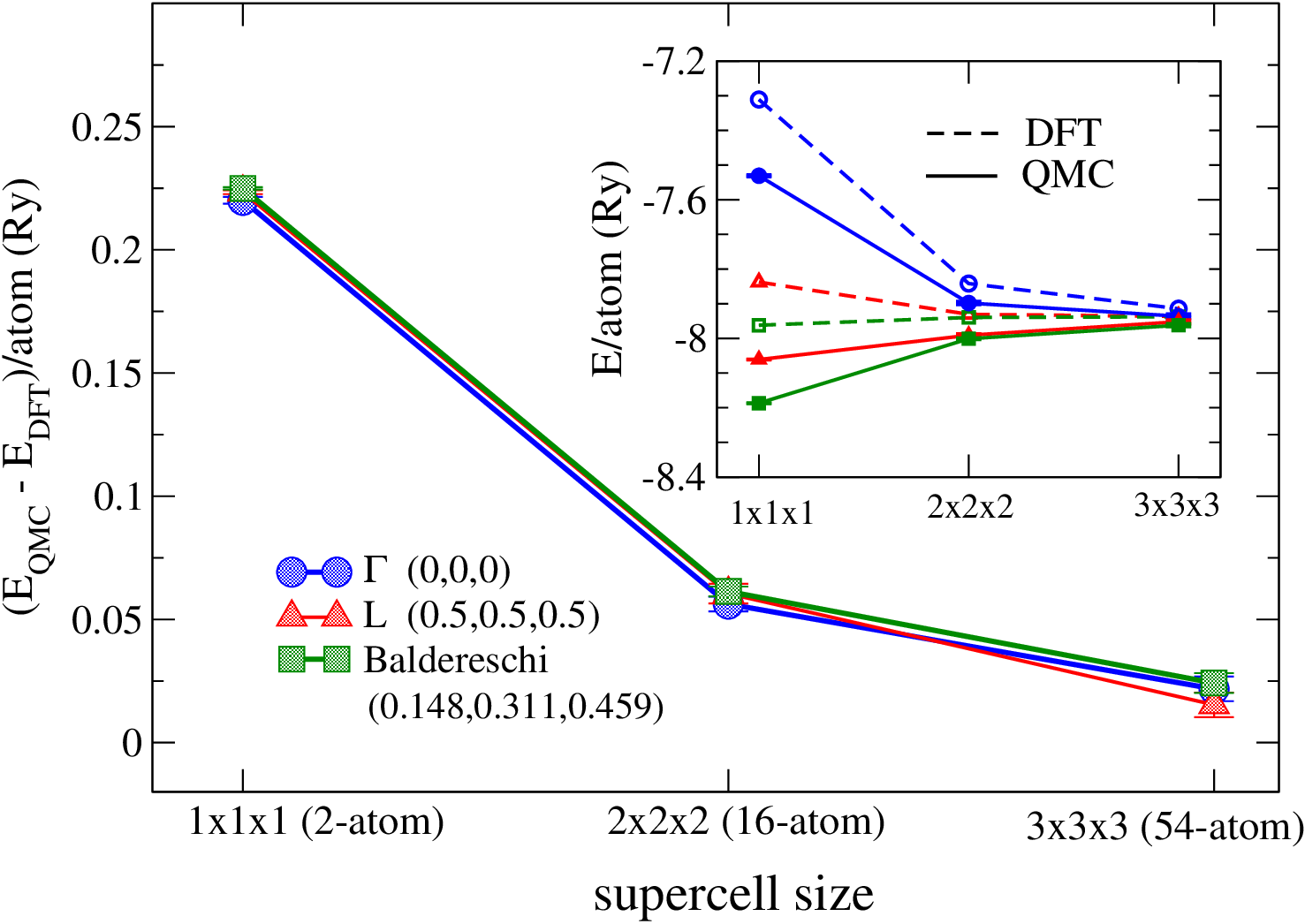}
\caption{(Color online)
Different behaviors of the FS errors in DFT and MB calculations, and the 
separation of the 1B and 2B FS errors.
The differences between the 
calculated AFQMC and DFT-LDA total energies in bulk silicon are shown in the main 
graph for 
 three special ${\mathbf k}$-points: $\Gamma$ (blue circle), $L$ (red triangle), and the Baldereschi \cite{Baldereschi-pt}  mean-value ${\mathbf k}$-point (green square). 
 The inset shows the calculated ground-state
energy. 
The dashed lines are DFT-LDA results, while the solid lines are corresponding AFQMC results.
 The supercell sizes are 
indicated on the horizontal axes.}
\label{fig:Si-QMC-DFT}
\end{figure}
The 2B FS error is the residual beyond the 1B FS error:
\begin{eqnarray}
\Delta E^{\rm 2B}({\mathbf k},L) &\equiv  & \Delta E({\mathbf k},L)-\Delta E^{\rm 1B}({\mathbf k},L) 
\nonumber \\ 
&=& E^\infty-E(\{{\mathbf k}\},L).
\label{eq:def_2B}
\end{eqnarray}
Thus $\Delta E^{\rm 2B}({\mathbf k},L)$ is independent of the ${\mathbf k}$ twist, {\it i.e.}, 
it is the FS error aside from ${\mathbf k}$-point sampling errors.

This is illustrated in Fig.~\ref{fig:Si-QMC-DFT},
which shows the total energy difference 
of bulk silicon between QMC and standard DFT LDA calculations. Three special 
${\mathbf k}$-points are used: 
$\Gamma$, $L$, and the Baldereschi \cite{Baldereschi-pt} 
mean-value ${\mathbf k}$-point. 
The 1B FS errors are seen to be rather parallel between the independent-electron and MB calculations.
The energy difference between QMC and DFT results,
the 2B FS error in QMC, 
is essentially independent of ${\mathbf k}$-points,
consistent with Eq.~(\ref{eq:def_2B}).
The 1B FS error in the MB calculation
can thus be removed by applying a correction $\Delta E^{\rm 1B}_{\rm DFT}({\mathbf k},L)$ from 
standard DFT. This is frequently done in practice. (Note that sometimes the 1B FS error has a different 
size dependence compared with that of the total $\Delta E({\mathbf k},L)$, 
as for the L-point in  Fig.~\ref{fig:Si-QMC-DFT}.
Applying this 1B FS correction would 
make the FS error {\it larger}.  \cite{kzk}) 
 A residual 2B FS error remains, however, which decays slowly with supercell size.

To consider the 2B FS error in the framework of DFT, specifically LSDA, we 
review how standard LSDA implicitly removes the 2B FS effect, leaving only 1B FS error,
which can be eliminated using sufficiently dense ${\mathbf k}$-point grids.
The many-electron Hamiltonian in a simulation cell is
(in Rydberg atomic units): 
\begin{equation}
H= -\sum_i \nabla_i^2 + \sum_i V_{\mathrm{ion}}({\mathbf r}_i) +
\sum_{i<j} V^{\rm FS}(|{\mathbf r}_i-{\mathbf r}_j|) \, ,
\label{eq:Hmb}
\end{equation}
where $i$ labels an electron, ${\mathbf r}_i$ is the electron position, 
the 
 ionic potential on $i$ can be local or non-local, and the sum runs over all electrons.
 We have suppressed the electron spin index, but spin is assumed to be present.                                                               
The Coulomb interaction $V^{\rm FS}$ between electrons depends on
the supercell size $L$ (and its shape),                                         
due to modification by the PBC,
for example, via
the Ewald method.  \cite{fraser}
The corresponding LSDA, as usually formulated, introduces a fictitious
mean-field system of electrons with the same external potential and the same 
spin configuration, with the single-electron Hamiltonian: \cite{hk,ks}                                    
\begin{equation}
H_{\rm DFT} = -\nabla^2 + V_{\mathrm{ion}}({\mathbf r}) + V_H({\mathbf r}) +
V_{xc}^\infty(n_\uparrow({\mathbf r}), n_\downarrow({\mathbf r}),{\mathbf r}) \, ,
\label{eq:KS}
\end{equation}
where the XC potential depends on the
electronic densities for the two spin species, and the Hartree term depends 
on the total density. The densities $n_\sigma({\mathbf r})$ are determined self-consistently,
with Fermi statistics imposed on the occupied orbitals (eigenfunctions of $H_{\rm DFT}$).                                    
The LSDA XC functional in Eq.~(\ref{eq:KS}) has an $\infty$ in the superscript: 
$V_{xc}^\infty$ is supposed to be the XC functional of the \emph{infinite} system. 
                       
The key is that
using $V_{xc}^\infty$ removes any 2B FS effect.
Typically
 $V_{xc}^\infty$ is obtained from QMC results on the homogeneous electron gas (HEG), 
extrapolated
to infinite size. \cite{ceperley78,ca,pz} Because 
LSDA yields an independent-electron problem, in which 
the Hamiltonian is only dependent on the electronic densities, Bloch's theorem can 
be applied. The only requirement for size convergence is the convergence of the densities,
which are fully defined by the periodicity of the system once the primitive cell is specified. 
The 
extrapolation of FS HEG data to the infinite size was important in that it eliminated the 
size dependence in the XC functional, making the theory consistent at the thermodynamic 
limit.

To have an LSDA theory for the FS supercell, we must recover the FS dependence of the 
interaction term, and make it parallel to the MB Hamiltonian:
\begin{equation}
H_{\rm DFT}^{\rm FS} = -\nabla^2 + V_{\mathrm{ion}}({\mathbf r}) + V_H({\mathbf r}) +
V_{xc}^{\rm FS}(n_\uparrow({\mathbf r}), n_\downarrow({\mathbf r}),{\mathbf r}) \, ,
\label{eq:FS-S}
\end{equation}
where the FS $V_{xc}^{\rm FS}$ represents the XC effects in the \emph{finite} simulation cell
in which the MB calculation is carried out. 
Within DFT, the XC energy functional $\varepsilon_{xc}^\infty$ is constructed based on the calculations of the HEG, and then applied to realistic system by the LSDA approximation. \cite{pw,bh,pz,ca,ceperley78}
Below, we examine the electron gas in finite
cubic supercells  (of linear size $L$) to obtain an ansatz for the FS XC energy
$\varepsilon_{xc}(L; n_\uparrow, n_\downarrow)$, from which 
we can derive the FS XC potential $V_{xc}^{\rm FS}$. A functional will be parametrized 
as a function of $L$.
With such a functional, LSDA calculations 
can be carried out in the same supercell as the MB calculation. The difference between 
the FS LSDA and the infinite LSDA results (e.g., from usual dense 
${\mathbf k}$-point calculations),  $\Delta E^{\rm FS}_{\rm DFT}({\mathbf k},L)$,
will provide an estimate of the MB FS error. The FS corrections used in the remainder of the paper can be summarized as:
\begin{equation}
\begin{array}{l}
  \Delta \mathrm{DFT}^{\mathrm{1B}} = E_ \mathrm{DFT}(\infty) - E_\mathrm{DFT}(L)\,, \\
 \Delta \mathrm{DFT}^{\mathrm{2B}} = E_\mathrm{DFT}(L) - E_\mathrm{DFT}^{\mathrm{FS}}(L)\,,\\
 \Delta \mathrm{DFT}^{\mathrm{FS}} =  \Delta \mathrm{DFT}^{\mathrm{1B}} + \Delta \mathrm{DFT}^{\mathrm{2B}}\,. 
 \end{array}
 \label{DeltaDFTFS}
 \end{equation}

\section{FS XC functional}
\label{sec:XC}

In this section we discuss the FS XC functional in the HEG. 
We will use $N$ to denote the total number of electrons in the supercell of volume 
$\Omega\equiv L^3$, and $N_\uparrow$
and $N_\downarrow$ for spin-$\uparrow$ and spin-$\downarrow$, respectively.
For convenience, we will refer to the majority spin as $\uparrow$-spin below, i.e., assuming
$N_\uparrow\ge N_\downarrow$.
The density is as usual specified by $r_s$, 
the radius of a sphere containing one electron on average, which is related to the 
charge density by $4\pi r_s^3 /3 \equiv 1/n$. 
The fractional spin polarization parameter $\zeta$ is defined as the ratio of spin density and charge density:
\begin{equation}
\zeta 
=\frac{n_{\uparrow}-n_{\downarrow}}{n_{\uparrow}+n_{\downarrow}},
\label{eq:zeta}
\end{equation}
where $n_{\uparrow}$ ($n_{\downarrow}$) is the electronic density of spin component up (down). The total charge density is $n$=$n_{\uparrow}$+$n_{\downarrow}$. The spin unpolarized state corresponds to $\zeta = 0$, while the fully spin polarized state to $\zeta = 1$.
The desired FS XC functional $\varepsilon_{xc}(L; n_\uparrow, n_\downarrow)$
is thus equivalently written as $\varepsilon_{xc}(r_s, L,\zeta)$ in the HEG.
The total XC energy is 
$\varepsilon_{xc}(r_s,L,\zeta)= \varepsilon_{x}(r_s,L,\zeta)+\varepsilon_{c}(r_s,L,\zeta)$. 
We discuss the exchange and correlation energies 
separately in Sec.'s \ref{subsec:exchange}
and \ref{subsec:correlation}  and summarize our parametrized functional with 
numerical parameter values in Sec.~\ref{subsec:xc-parameters}.

\subsection{Exchange energy functional}
\label{subsec:exchange}
 
The exact total energy of the HEG can be separated into the HF energy, which is the sum of kinetic energy and the exchange energy, and the remainder, which is termed  correlation energy. The kinetic energy is simply given by a sum of the energies of independent electrons filled from the lowest single-particle energy state up to the Fermi surface. 
The exchange energy from the filled Fermi sphere can be calculated analytically, 
and is of the form $\propto 1/r_s$.
The asymptotic expression of the 
correlation energy in the high density region was derived by Gell-Mann and Brueckner, \cite{gm} which is used in combination with 
DMC results at various densities \cite{ca} to obtain a parametrized fit \cite{pz} of the LSDA 
XC energy.
As mentioned, all of these are at the infinite size limit.

In finite-sized supercells, the HF exchange energy of the HEG 
scales as $1/L$.
The FS error in the  exchange energy per electron, 
$\Delta \varepsilon_x(r_s,L) \equiv \varepsilon_{x}^{\infty}(r_s) - \varepsilon_{x}(r_s,L)$, 
scales \cite{kzk} as $1/r_s$,
with a coefficient 
$\Delta \upsilon(N)$ which is only dependent on $N$ (i.e., the ratio
of $L/r_s$),  not on $r_s$. 
The FS exchange energy approaches the  infinite size limit from below.
$\Delta \upsilon(N)$ decays smoothly like $1/N^{2/3}$ when averaged over many 
${\mathbf k}$-points. \cite{ceperley78} Thus the first order correction of exchange energy will be of order $r_s/L^2$. We will include a next-order term in the parametrization, $r_s^2/L^3$, which follows the same scaling relation.

In a system with two spin components, 
each component contributes independently to the total exchange energy. There is no exchange interaction between opposite spins. 
Thus, for a system of polarization $\zeta$, 
the exchange energy is given by
\begin{equation}
\varepsilon_x(r_s,\zeta)=\frac{1+\zeta}{2}\varepsilon_x(r_{s\uparrow},0)+\frac{1-\zeta}{2}\varepsilon_x(r_{s\downarrow},0)
\label{eq:Ex}
\end{equation}
where $\varepsilon_x(r_s,0)$ is the exchange energy of the unpolarized system with total density given by $r_s$, and the density for the  spin up (down) component is given by 
$r_{s\uparrow}=r_s (1+\zeta)^{-1/3}$ ($r_{s\downarrow}=r_s (1-\zeta)^{-1/3}$). 
With this relation, we can obtain the FS exchange energy for an arbitrary spin polarization.

However, the FS exchange energy for spin unpolarized system is discontinuous as given
by KZK. \cite{kzk} At low densities, the 
exchange energy is forced to go to zero rapidly, because in the \emph{finite} supercell 
there will only be less than one electron in each spin component 
when $r_s$ is larger than some threshold $\gamma_x$. In this regime, the functional form 
from scaling, i.e., $r_s^{n-1}/L^n$, would diverge. A different functional form is used instead, 
and the two are joined at $r_s=\gamma_x$ by requiring the exchange potential to be continuous,
which resulted in a discontinuity in the exchange energy. 
With this form of the exchange energy, the use of Eq.~(\ref{eq:Ex}) would lead to multiple 
discontinuities in a partially polarized system, and an exchange that tends to be too high 
in the intermediate densities which connect the two density regimes. 
 
 Instead of the formula in  Eq.~(\ref{eq:Ex}), we will make an interpolation between spin unpolarized and fully polarized systems. The interpolation formula \cite{bh,pz} is
\begin{equation}
\varepsilon_{x}(r_s,L,\zeta)=\varepsilon_{x}(r_s,L,0)+f(\zeta)[\varepsilon_{x}(r_s,L,1)-\varepsilon_{x}(r_s,L,0)],
\label{eq:EX-interpolation}
\end{equation}
with the function $f(\zeta)$ defined as
\begin{equation}
f(\zeta)=\frac{(1+\zeta)^{4/3}+(1-\zeta)^{4/3}-2}{2^{4/3}-2}.
\label{eq:fzeta}
\end{equation}
The formula in Eq.~(\ref{eq:EX-interpolation}) gives the exchange energy for any polarization 
by interpolation bwteeen the exchange energies for unpolarized and fully polarized systems. 
When $L$ is infinite, this formula reduces to the exact relation in Eq.~(\ref{eq:Ex}).
The use of this formula also makes the treatment of the exchange consistent with that 
of the correlation (see Sec.~\ref{subsec:correlation}), 
for which there is no analogy of  Eq.~(\ref{eq:Ex}).

We will parametrize the exchange energies for the two end point systems
in the following form:
\begin{equation}
\varepsilon_x(r_s,L,p)=
\begin{cases}
     \frac{a_0(p)}{r_s} + \frac{a_1(p)}{L^2}r_s +  \frac{a_2(p)}{L^3}r_s^2\,, & \text{if} \  r_s \le \gamma_x\,; \cr \\
     \frac{a_3(p) L^5}{r_s^6}+\frac{a_4(p) L^6}{r_s^7} +\frac{a_5(p) L^7}{r_s^8}\,,  & \text{other}.\cr
\end{cases}
\label{eq:Exnew}
\end{equation}
Here $p$ has two discreet values: $p=0$ for spin unpolarized while $p=1$ for fully polarized. 
The formula is the same as that in Ref.~\onlinecite{kzk} for the high density region, but 
we have introduced two additional terms in the low density part. 
The values of $a_0(p)$ are determined by the infinite size HEG exchange energies.
The other coefficients are obtained from a fit and by continuity conditions 
at $\gamma_x$. As described in Sec.~\ref{subsec:xc-parameters}, we choose the same 
$\gamma_x$ for $p=0$ and $p=1$, hence for all polarizations $\zeta$. 
At $r_s=\gamma_x$, we require the exchange potential and its first derivative to be 
continuous for all polarizations. The conditions are straightforward to impose at 
$p=0$ and $p=1$.  For an arbitrary spin polarization, the spin interpolation function $f(\zeta)$ 
in Eq.~(\ref{eq:EX-interpolation}) depends on the spin density. The exchange potential is given by
\begin{align}
&\nu_{x,\uparrow(\downarrow)}(r_s,L,\zeta) =\frac{\partial (n\,\varepsilon_{x}(r_s,L,\zeta))}{\partial n_{\uparrow(\downarrow)}}  \nonumber \\
&=\nu_{x}(r_s,L,0)+f(\zeta)[\nu_{x}(r_s,L,1)-\nu_{x}(r_s,L,0)] \nonumber \\
&+n\frac{\partial f(\zeta)}{\partial n_{\uparrow(\downarrow)}} [\varepsilon_{x}(r_s,L,1)-\varepsilon_{x}(r_s,L,0)].
\label{eq:Vx}
\end{align}
We see that, if the energy difference between the polarized and unpolarized systems, $\varepsilon_{x}(r_s,L,1)$-$\varepsilon_{x}(r_s,L,0)$,
is made continuous at $r_s=\gamma_x$, the exchange potential will be continuous for 
any $\zeta$. 
It is easy to show that continuity of the first derivative of the exchange potential is also satisfied with these conditions. 
We calculate the exchange energies for different supercell sizes  
by averaging over ${\mathbf k}$-points. The quality of the fits is consistent with that in
Ref.~\onlinecite{kzk}. 
Numerical values of the parameters are given in Sec.~\ref{subsec:xc-parameters}.

\subsection{Correlation energy functional}
\label{subsec:correlation}

Unlike the exchange energy, the correlation energy consists of both intra- and inter-spin contributions. There is no exact mapping of the correlation energy at an arbitrary polarization 
from spin unpolarized results. 
Within standard DFT, the correlation energy for partial spin polarization is obtained from an interpolation between spin unpolarized and fully spin polarized systems, as we have done 
with the exchange energy. We will follow the same procedure. Below, we first
construct the energy functions for the two end-point systems of $p=0$ and $p=1$.

In Ref.~\onlinecite{kzk}, the correlation energy of unpolarized system of the HEG is expressed separately for high, intermediate, and low density regions. In the high density region, the formula is parameterized by fitting to Ceperley and Alder's extrapolation fit \cite{ca,ca87,tanatar,kwon}
\begin{flalign}
\begin{split}
\quad \varepsilon_c(r_s,L)& = \varepsilon_c^{\infty}(r_s)-\Delta \varepsilon_x(r_s,L)  \\
& +[b_1(r_s)-1]\Delta \varepsilon_K(r_s,L) + b_2(r_s)/N\,,
\end{split}
\label{eq:extrapolation}
\end{flalign}
where the parameters $b_1(r_s)$ and $b_2(r_s)$ were written as low order polynomials
in $r_s^{1/2}$ and $r_s^{-1/2}$, respectively, and fitted to tabulated data for unpolarized 
systems. \cite{kwon,hendra} 
(Note that here $\Delta \varepsilon_x(r_s,L)$ and $\Delta \varepsilon_K(r_s,L)$ are given by $\varepsilon_x(r_s,L)$-$\varepsilon_x^{\infty}(r_s)$ and $\varepsilon_K(r_s,L)$-$\varepsilon_K^{\infty}(r_s)$, respectively.)
To avoid the divergence at sufficiently large $r_s$ or small $N$, the correlation energy is set to zero. In the intermediate density region, 
a polynomial function is used whose parameters were completely determined from continuity conditions. 
For fully polarized systems, the available finite-size data from QMC were not as
detailed. We instead modify the procedure above, by generating FS HEG data in the intermediate 
density regime and fitting them to obtain a unified function extending to higher densities.

\begin{figure}
\includegraphics[width=0.48\textwidth]{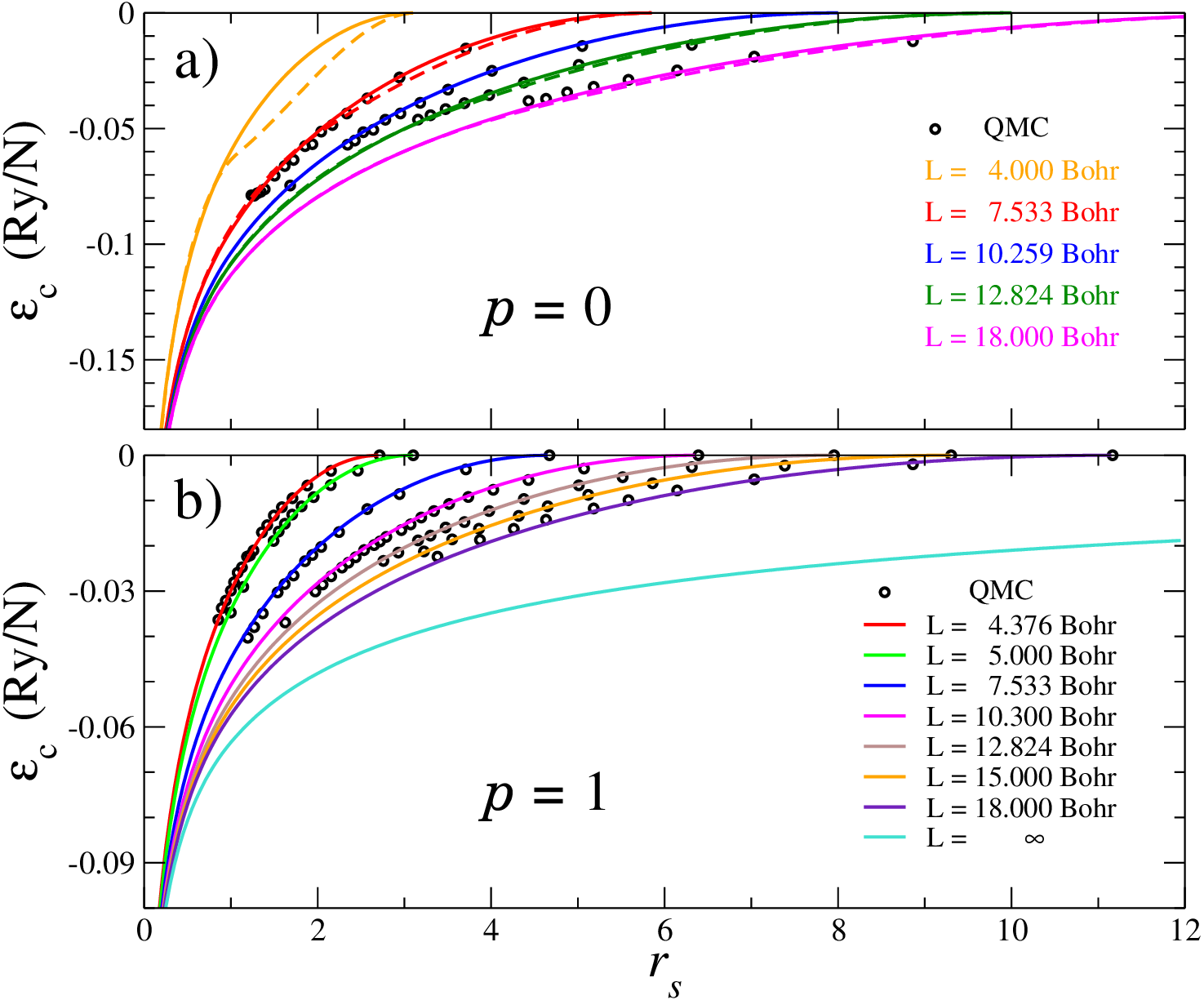}
\caption{(Color online)
Quality of the new FS correlation functional.  
a) The new parametrization of the correlation energy functional for the spin unpolarized HEG is 
compared with AFQMC data and with the previous functional of KZK. \cite{kzk}
The solid lines are the new form given by Eq.~(\ref{eq:Ec}), while the dashed lines 
are from KZK. 
b) The parametrized correlation energy for  fully spin polarized HEG systems is
compared with AFQMC data. Solid lines represent the fits while symbols are from AFQMC 
at finite sizes. Statistical error bars in the AFQMC results are smaller than the symbol size. 
The infinite size LSDA result is also shown for $p=1$. 
}
\label{fig:Fit}
\end{figure}

We parameterize the spin unpolarized and fully spin polarized correlation energies by a new unified function with only two density regions:
\begin{equation}
\varepsilon_{c}(r_s,L,p)=
\begin{cases}
     \varepsilon_c^{\infty}(r_s,p)-\frac{a_1(p)}{L^2}r_s+\frac{g(r_s,L,p)}{L^3}, &
     \text{if}\ r_s \le \gamma_c;\cr \\
      0, &     \text{other}.\cr
\end{cases}
\label{eq:Ec}
\end{equation}
The functional form is chosen to approach the correct asymptotic value at high density, where
we have used the Perdew-Zunger (PZ) parametrization \cite {pz} for the 
function $\varepsilon_c^{\infty}(r_s, p)$. 
The leading FS term $a_1(p)$  exactly cancels its counterpart in the exchange energy 
to ensure that the total XC energy 
scales as ${\mathcal O}(1/L^3)$ as expected. For the $1/L^3$ term, we choose 
the functional form
\begin{flalign}
\begin{split}
 g(r_s,L,p)&\equiv g_1(L,p)\,r_s \ln(r_s) + g_2(L,p)\,r_s  + g_3(p)\,r_s^{1/2} \\
 &+ g_4(p)\,r_s^{3/2} \ln(r_s) + g_5(p)\,r_s^{3/2} + g_6(p)\,r_s^2.
\end{split}
 \label{eq:gr_Ec}
\end{flalign}
The last four parameters are determined by the fits, while 
the other two 
are determined
by continuity conditions. We require the correlation energy and its first derivative to be continuous
at $\gamma_c(p)$. 
Our choice of $\gamma_c(p)$
 and the final parameter values are given in Sec.~\ref{subsec:xc-parameters}.

 We perform phaseless AFQMC \cite{afqmc} calculations of the 
 correlation energies of the HEG, for $p=1$ and $p=0$, 
 in finite supercells at typical densities. With each choice of $N$
 and $L$, we average the total energy over multiple ${\mathbf k}$-points. The results are shown 
 in Fig.~\ref{fig:Fit}. it is seen that the new parametrized functions fit all 
 the AFQMC data well. Figure~\ref{fig:Fit}a also shows a comparison between the 
 new functional and that of KZK for unpolarized systems. The two fits are almost  
 indistinguishable, except in the intermediate region, where the new function shows a slightly better fit to AFQMC results, especially for small lattice sizes. As mentioned, in the approach of KZK, the functional form at high density is 
 fully determined by existing DMC data. AFQMC data were only used to guide the choice 
 of the boundaries which separated the high density regime from the intermediate and 
 low density regimes. In contrast, the current approach only had AFQMC data at intermediate densities
 to determine the functional form for $r_s\le \gamma_c$. 
 The good agreement between the 
 two parametrizations across the density range is thus very reassuring. 
  Contrary to the exchange energy, the correlation energy approaches the thermodynamic limit from above, as illustrated in Fig.~\ref{fig:Fit}b.

\begin{figure}
\includegraphics[width=0.48\textwidth]{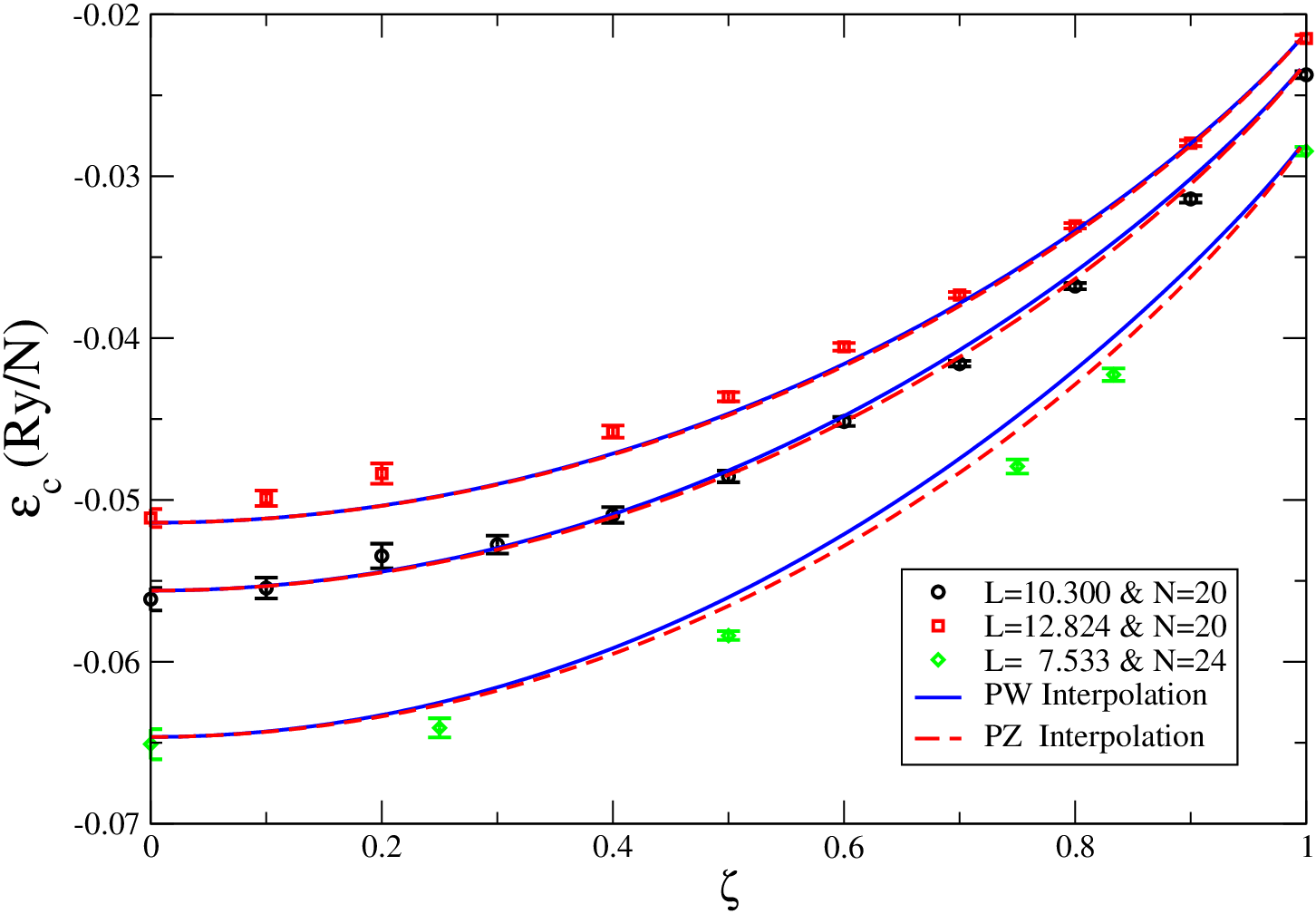}
\caption{(Color online)
Interpolations of the FS correlation energy for partial spin polarizations in the HEG.
Three supercell sizes are shown, each with two lines showing the two interpolation schemes.
Corresponding FS AFQMC results are shown together with statistical error bars.
}
\label{fig:Ec-Inter}
\end{figure}

We have now obtained the FS correlation energy formulas for spin unpolarized ($p=0$) and fully spin polarized ($p=1$) systems, with which we can interpolate to arbitrary spin polarization $\zeta$.
As mentioned, the interpolation of the correlation energy is not exact. 
In standard DFT, two forms have been widely adopted: \cite{bh,pw,pz,ob} that of PZ as we have 
used for the exchange in Eq.~(\ref{eq:EX-interpolation}), and that of Perdew and Wang (PW): 
\begin{eqnarray}
\varepsilon_{c}(r_s,L,\zeta)=\varepsilon_{c}(r_s,L,0)+\alpha_c(r_s) \frac{f(\zeta)}{f^{''}(0)} [1-\zeta^4] \nonumber \\
+ [\varepsilon_{c}(r_s,L,1)-\varepsilon_{c}(r_s,L,0)] f(\zeta) \zeta^4,
\label{eq:PW-interpolation}
\end{eqnarray}
where $\alpha_c(r_s)$ is the spin stiffness function
which includes additional fitting parameters. The interpolations for several different supercell sizes are shown in Fig.~\ref{fig:Ec-Inter}. 
In the PW results, we have fixed the $\alpha_c(r_s)$ at the infinite-size values without further 
parametrization. It is seen that the two schemes give similar results. Compared to the 
 actual FS data from AFQMC calculations in polarized systems, both schemes provide reasonable 
 estimates. Discrepancies are visible, especially at smaller supercell sizes. 
In the applications below, we have used the PZ interpolation scheme for the correlation
energy, due to its simplicity, {\it i.e.}, using the same interpolation as in the exchange energy in Eq.~(\ref{eq:EX-interpolation}).
The PZ form is equivalent to the special choice \mbox{$\alpha_c(r_s)=f^{''}(0)[\varepsilon_{c}(r_s,L,1)-\varepsilon_{c}(r_s,L,0)]$}
in Eq.~\ref{eq:PW-interpolation}.
 This is 
 further discussed in Sec.~\ref{sec:discuss}.

\subsection{Numerical Parameters}
\label{subsec:xc-parameters}

\begin{table}
\caption{\label{tab:xc-table} Coefficients in the parametrized FS LSDA XC functionals for spin unpolarized and fully polarized systems. All values are given in Rydberg atomic units. The choices of the continuity densities $\gamma_x$ and $\gamma_c$ are given in Sec. \ref{subsec:xc-parameters}.}
\newcolumntype{d}[1]{D{.}{.}{#1}}
\begin{tabular}{|c|d{5}|d{5}|d{5}|d{5}|}
\hline
 \multirow{2}*{\itshape i} & \multicolumn{2}{c|}{\itshape p=0} &\multicolumn{2}{c|}{\itshape p=1 } \\
\cline{2-5}
   & \multicolumn{1}{c|}{$a_{i}(0)$} & \multicolumn{1}{c|}{$g_{i}(0)$}  & \multicolumn{1}{c|}{$a_{i}(1)$}   &  \multicolumn{1}{c|}{$g_{i}(1)$} \\
\hline
0  & -0.9163  &           & -1.1545   &          \\
1  & -2.2037  &           & -1.7491   &          \\
2  &  0.4710  &           &  0.2967   &          \\
3  &  0.2339  &   0.2109  &  0.1812   &  0.7528  \\
4  & -0.4880  &   8.4987  & -0.4515   &  3.3314  \\
5  &  0.1847  & -13.6840  &  0.1786   & -5.1050  \\
6  &          &  -4.6977  &           & -2.3048  \\
\hline
\end{tabular}
\end{table}

The functional forms of our parametrized FS LSDA XC functional are given in 
Eqs.~(\ref{eq:Exnew}), (\ref{eq:Ec}), and (\ref{eq:gr_Ec}).
By performing AFQMC calculations and fitting to the results with the procedures 
described above, we obtain the values of the 
coefficients, which are given in Table~\ref{tab:xc-table}.
The only remaining parameters are the densities at which the different pieces in the
functional are joined. 
For the exchange energy, we used 
 $\gamma_x= r_s(N=1)$ for both spin unpolarized and fully polarized states.
 For the correlation energy, we chose
 $\gamma_c= r_s(N=1)$ for the fully polarized system and
 $\gamma_c= r_s(N=0.5)$ for the unpolarized system.
 
These choices are guided by the basic idea that beyond $\gamma$, we have 
a supercell of the HEG in which there is only a fraction of an electron. The exchange (correlation) 
energy needs to vanish rapidly in this regime. Clearly the location
of $\gamma$ cannot be uniquely defined. The results from the XC functional 
should not be sensitive to the precise value of $\gamma$. We have verified that this is indeed 
the case in typical supercell sizes. 
In principle, 
$\gamma_x$ should depend on the spin polarization $\zeta$. For the fully polarized system,
$N=1$ signals the transition point, while for the unpolarized system, $N=2$. We have taken 
them to be the same value, which simplifies the interpolation and makes the functional 
easier to implement, because all values of $\zeta$ share the same $\gamma_x$.
The form of the correlation energy is continuous and it is easier to have different  $\gamma$ 
values for different $p$. 
We have thus chosen $\gamma_c$ for $p=0$ to be the same
as in KZK for consistency.

In the correlation energy, there are six parameters in the formula of Eq.~(\ref{eq:gr_Ec}).
We only need to fit the last four, since the other two, $g_1$ and $g_2$, are fully 
determined by the continuity conditions. This is why they have a dependence on lattice size $L$.
In the applications below, we have used the PZ interpolation scheme in the correlation
energy, i.e., using the same interpolation as in the exchange energy. 
The $1/L^{2}$ term is then exactly canceled at any polarization, 
leaving only $\mathrm{1/\Omega}$  and
higher order terms in the FS errors.

\section{Applications}
\label{sec:apps}

We consider four examples here to illustrate the application of the 
new FS LSDA functional. The four problems are chosen to cover different aspects of the application. 
All of them have spin polarizations or magnetic ground states which require 
the new spin-dependent FS functional.
Calculations in single atoms and molecules provide systematic MB results for arbitrary 
supercell sizes which allow a detailed study of the FS convergence. The ionization energy 
of Mn is calculated to study the effect of the FS LSDA in supercells when there is a physical charge 
imbalance. In the last example, transition metal oxide systems MnO and FeO are considered to study the application of the FS correction in extended 
systems with magnetic order.
In the first three examples, we use the phaseless AFQMC method  \cite{afqmc,malliga} in a 
plane-wave basis, with a norm-conserving pseudopotential to carry out the MB calculations,
before applying the FS corrections. More systematic AFQMC calculations of 
atomic and molecular systems using plane-wave basis exist.  \cite{malliga,afqmc} Our goal 
here is simply to use it as an accurate MB approach for model extended systems to study 
FS corrections.  
In the last example, we use existing DMC results \cite{mno,feo} and use the 
new FS functional as a post-processing correction.

\begin{figure}
\includegraphics[width=0.48\textwidth]{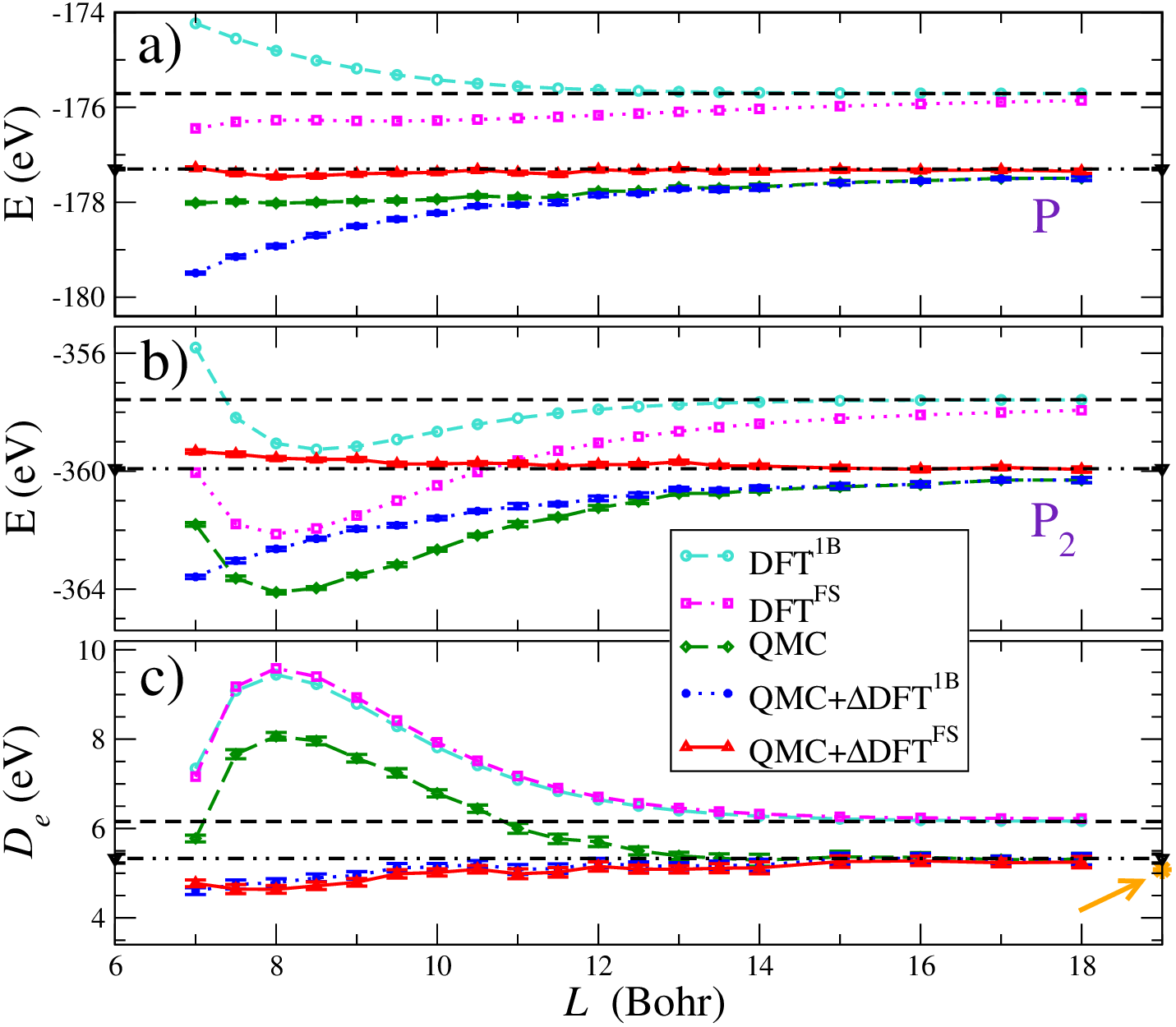}
\caption{(Color online)
Convergence of the P$_2$ molecular dissociation energy vs. supercell size. 
Calculated AFQMC total energies of the P atom and P$_2$ molecule are shown in panels (a) and (b), respectively.
The corresponding $D_e$~of P$_2$ is shown in (c). 
The infinite-size limits of DFT and AFQMC are shown by the
dashed and 
dot-dashed lines, respectively. 
AFQMC statistical error bars
are shown but many are smaller than the symbol size. The orange arrow in (c) 
points to the experimental value.
}
\label{fig:P}
\end{figure}

Our first application is to calculate the dissociation energy ($D_e$) of the P$_2$ molecule, using a 
plane-wave basis and
PBC 
with single ${\mathbf k}$-point sampling, \mbox{${\mathbf k}=\Gamma$}. 
In both AFQMC and DFT calculations, a norm-conserving Kleinman-Bylander \cite{kb} separable nonlocal LDA pseudopotential \cite{opium} is used.
The P atomic ground state is magnetic, with an electronic configuration 
\mbox{$3s[\uparrow\downarrow]\,3p[\uparrow\uparrow\uparrow]$}, a ``close-shell" system for each spin. The P$_2$ molecular ground state is non-magnetic, with 5 $\uparrow$-spin and 
 5 $\downarrow$-spin electrons.
 The total energy calculations are performed in cubic supercells of size $L = 7 \mathrm{-} 18$\,Bohr. 
 Figure~\ref{fig:P} shows the MB results from AFQMC calculations, as well as 
 DFT results from both the standard, infinite-size LSDA XC functional (denoted by DFT$^\infty$) and the new FS functional (denoted by DFT$^{\mathrm{FS}}$). 
 
 The conventional DFT$^\infty$ energies converge to the infinite limit very rapidly, as expected.
 The DFT$^{\mathrm{FS}}$ calculations exhibit a significantly slower convergence, 
 reflecting the 2B FS effect which it is designed to capture. 
 For the P atom, the overall FS effect is not very large, but a persistent FS error can be seen 
 in the raw AFQMC results.  Up to $L = 18$\,Bohr, the energy is still about $0.15$\,eV away from the infinite-size value.
 The 1B correction with standard DFT 
  [$\Delta \mathrm{DFT}^{\mathrm{1B}}$; see Eq.~(\ref{DeltaDFTFS})] is in the wrong direction and leads to a
 larger FS error and slower rate of convergence. 
 By contrast, the DFT$^{\mathrm{FS}}$ total energies, which include both 1B and 2B corrections
  [Eq.~(\ref{DeltaDFTFS})],  show size dependence similar to the AFQMC results. 
After $\Delta \mathrm{DFT}^{\mathrm{FS}}$ correction, 
the AFQMC total energy 
 converges to the asymptotic value rapidly, as seen from the 
 essentially flat curve beyond $L=9$\,Bohr.
 
 A similar trend is seen in the P$_2$ molecule as shown in Fig.~\ref{fig:P}b.  In this case the 1B correction does reduce the FS error, but only up to intermediate size $L$, beyond which 
 the standard DFT has reached convergence. 
 The MB results remain unchanged for large $L$, where the 2B FS effect dominates. The DFT$^{\mathrm{FS}}$ calculation again tracks the FS dependence of the MB AFQMC, and the total energy of the latter converges very rapidly to the infinite size limit after applying the full FS correction. The P$_2$ system is unpolarized, so the KZK parametrized FS 
 functional can also be applied. Compared to the results there, \cite{kzk} 
 the new functional seems to slightly overestimate the FS error. This small difference is 
 from the choice of exchange energy discontinuity point $\gamma_x$, as mentioned in 
 Sec.~\ref{subsec:xc-parameters}. 
With the FS corrections, the calculated $D_e$~shows good convergence by about $L\sim12$\,Bohr,
as seen in Fig.~\ref{fig:P}c.

\begin{figure}
\includegraphics[width=0.48\textwidth]{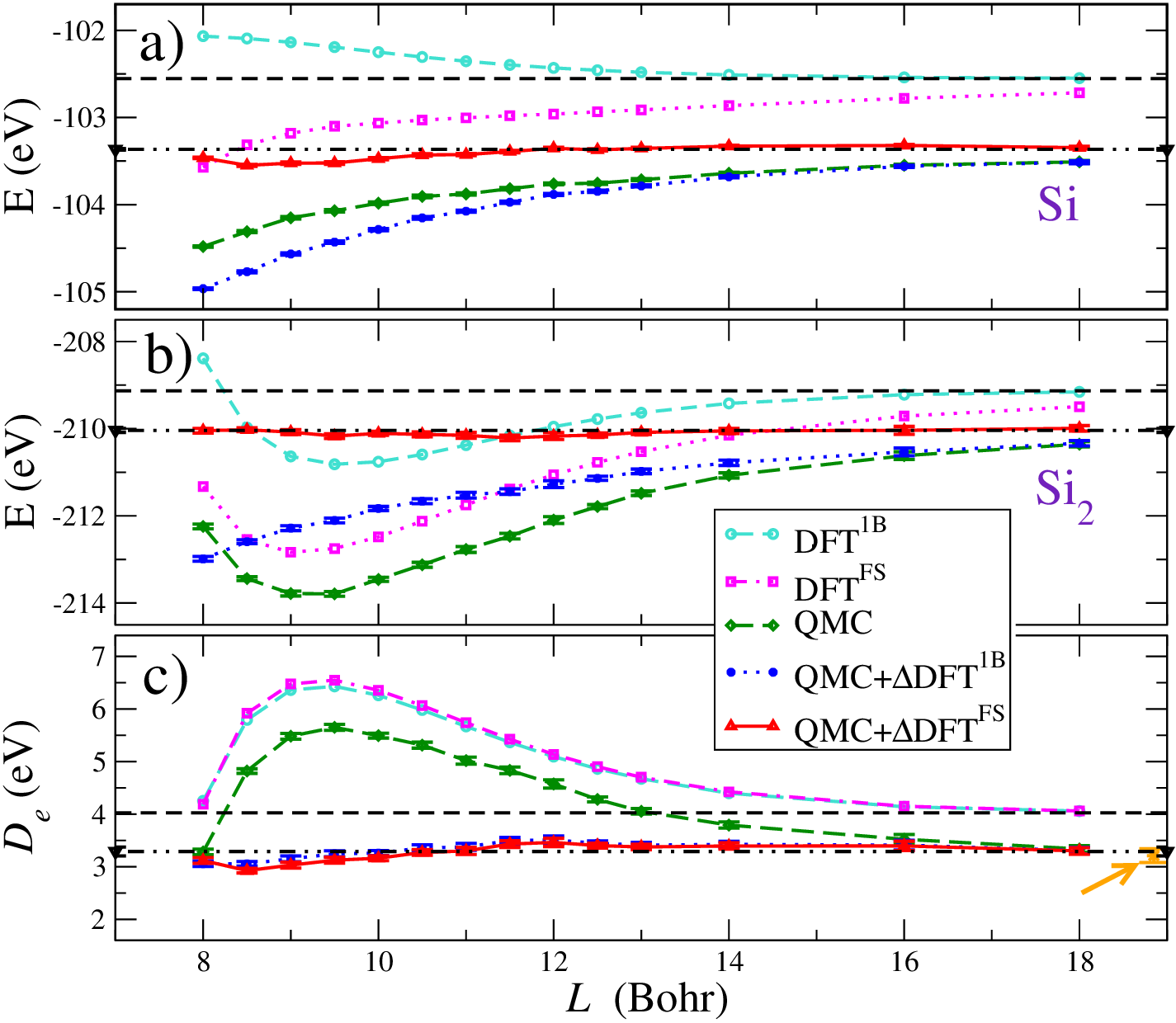}
\caption{(Color online)
Convergence of the Si$_2$ molecular dissociation energy vs. supercell size.
Same conventions as in Fig.~\ref{fig:P}.
 }
\label{fig:Si}
\end{figure}

Our second application is for the Si atom and molecule. These are ``open-shell" systems, and both 
systems are spin polarized in the ground state
($3\uparrow1\downarrow$ and $5\uparrow3\downarrow$, respectively).
The calculations are done in a similar way to the first application, with planewave basis and a
norm-conserving non-local pseudopotential in a periodic supercell.
The dependence of total energies and the dissociation energy with lattice size $L$ is presented in Fig.~\ref{fig:Si}.
Similar to the P atom, the usual DFT 1B correction for the Si atom has an opposite sign from the total FS error. With DFT$^{\mathrm{FS}}$ corrections, the convergence of the 
total energy of the atom is greatly improved across the entire range of lattice size studied. 
Similarly, for the 
Si$_2$ molecule, the total energy becomes essentially flat when the full 
correction is included, leading to rapid convergence to the infinite-size limit.

Although the atoms and molecules are relatively simple systems for MB calculations, 
they provide excellent tests for the FS correction method. They are model systems of 
a molecular solid in which the supercell size can be arbitrarily and systematically varied. 
Because of the nature of the interactions and the highly inhomogeneous density distributions
in the supercell, LSDA calculations are not particularly accurate, as can be seen 
by the final DFT $D_e$'s (panel (c) in Figs.~\ref{fig:P} and \ref{fig:Si}). Yet the FS correction method 
based on LSDA works extremely well. This underscores an important point, namely that
for the FS correction method to be effective, the system does not \emph{necessarily} have to be 
weakly or moderately correlated. 
The FS correction method requires LSDA 
to provide a good approximation in capturing the {\em difference\/}
between the system with interaction $V^{\rm FS}$ and that with 
the full interaction $V$ (no PBC image interactions). \cite{kzk}
This is not the
same as requiring LSDA to work well in either system. 
As the supercell size increases, the correction decreases and approaches zero. 
One would expect the correction to be ineffective if the supercell 
is less than the size 
of the XC hole, which would cause a distortion that is not captured in any LSDA approximation.

\begin{figure}
\includegraphics[width=0.48\textwidth]{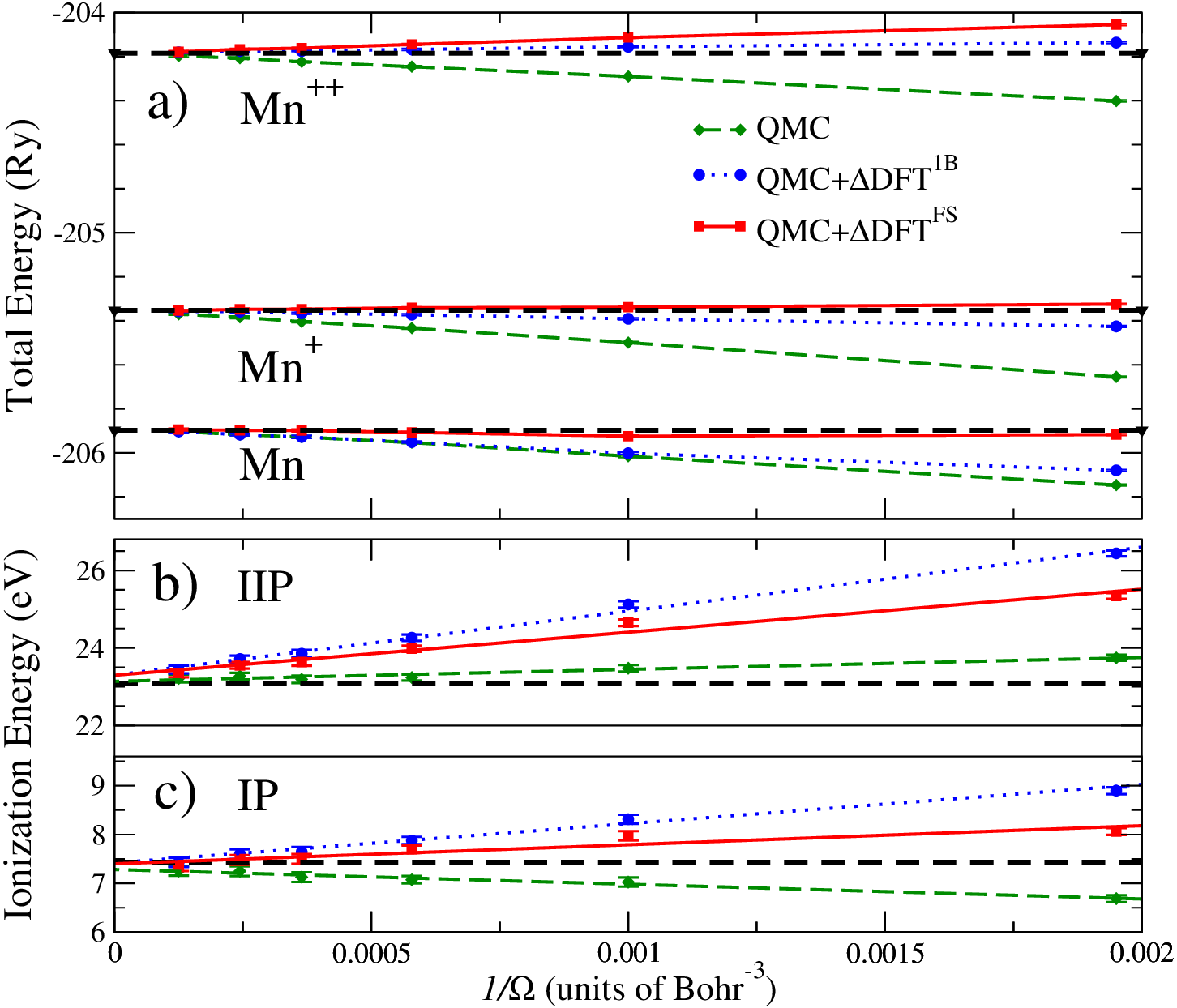}
\caption{(Color online)
Convergence of the 
ionization energies of the Mn atom vs.~supercell size
(plotted as the inverse of the supercell volume 1/$\Omega$).
Panel (a) shows the calculated AFQMC total energies for Mn, Mn$^{+}$, and Mn$^{++}$; the corresponding first (IP) and second (IIP) ionization energies are
shown in panels (b) and (c). Curves labeled QMC  are from raw AFQMC results [for the charged ions, the leading order $\Delta\mathrm{MP}^{(1)}$ correction is included (see text)]. The 
$\Delta\mathrm{DFT}^{\mathrm{1B}}$ and $\Delta\mathrm{DFT}^{\mathrm{FS}}$ corrections are as in Figs.~\ref{fig:P} and \ref{fig:Si} [for the charged ions, these corrections also add the $\Delta\mathrm{MP}^{(2)}$ correction (see text).]
Statistical error bars 
are shown and are about the size of the symbols.
 The heavy black dash lines indicate the infinite size limits of the calculated results in panel (a) 
 and the experimental ionization energies in panels (b) and (c).
}
\label{fig:Mn}
\end{figure}

The third application is to 
the Mn transition metal atom and its first and second ionization potentials. The ${\mathbf k}$-point used here is (0.25,0.25,0.25). 
For charged systems such as Mn$^+$ or Mn$^{++}$, there is an additional 
complication in a supercell calculation, since charged systems are incompatible with PBC.
A uniform neutralizing background is introduced to maintain charge neutrality, \cite{mp} 
but this results in a slow ${\mathcal O}(1/L)$ size convergence  to leading order.\cite{ewald,ziman} 
In independent-electron calculations such as DFT, Makov and Payne (MP) \cite{mp,pwscg} argued that the dependence has
the form:
\begin{equation}
E(L)=E(\infty)-\frac{q^2\alpha}{2 L}+\frac{2\pi qQ}{3 L^3}+\mathcal O(L^{-5}),
\label{eq:mp}
\end{equation}
where $q$ is the neutralizing charge, 
$\alpha$ is the Madelung constant,
and $Q$ is the quadrupole moment of the supercell. 
Thus, adding the counter-terms $\Delta$MP$^{(1)} \equiv \frac{q^2\alpha}{2 L}$ 
and $\Delta$MP$^{(2)}\equiv -\frac{2\pi qQ}{3 L^3}$ to $E(L)$ accelerates the size-convergence of
DFT calculations for charged systems.
The $\Delta$MP$^{(1)}$ correction applies equally well to a many-body calculation with PBC, since it depends only on the neutralizing charge, 
and all AFQMC results reported below for charged systems include this important correction. The $\Delta$MP$^{(2)}$ correction
depends on $Q$, which will generally differ, in a many-body calculation, from the DFT value.
In this work we have not calculated $Q$ 
in AFQMC, but have used the DFT value.

Figure \ref{fig:Mn} 
shows the size convergence of total energies of Mn, Mn$^{+}$, and Mn$^{++}$ and the corresponding first and second ionization energies.
For Mn and Mn$^{+}$ total energies, $\Delta\mathrm{DFT}^{\mathrm{FS}}$ removes nearly all of the FS error, while it somewhat overcorrects 
for Mn$^{++}$. The $\Delta\mathrm{DFT}^{\mathrm{1B}}$ corrections are smaller in all cases. 
For the ionization energies, both $\Delta\mathrm{DFT}^{\mathrm{1B}}$  and
 $\Delta\mathrm{DFT}^{\mathrm{FS}}$ are seen to overcorrect the FS error. For IP, the residual error after
 $\Delta\mathrm{DFT}^{\mathrm{FS}}$ correction is slightly smaller than the raw AFQMC result, 
 while $\Delta\mathrm{DFT}^{\mathrm{1B}}$ increases the FS error. For IIP, both FS corrections increase the FS error.
Extrapolation of these results to the infinite size limit, 
leads to IP and IIP values of 7.27(08) and 23.15(08) eV, respectively. These are 
in reasonably good agreement with 
the experimental values of 7.43 eV and 23.07 eV, respectively. \cite{sugar}
Because of strong FS error cancellations, the raw QMC IP and especially IIP energies show
little FS effect. This magnifies the residual errors in the FS correction for charged systems.
The most likely source of the residual error would appear to be the discrepancy in applying
the $\Delta$MP$^{(2)}$ directly from DFT. The MB and DFT estimates of the 
quadrupole moment $Q$ are likely to differ. We will leave further investigation of this effect to 
a future study. The total energies, after FS correction, show much smaller FS error, especially
in the neutral system of Mn. Given the strong magnetic nature of this 
 transition metal system, this is an encouraging result for our FS correction scheme.

\begin{figure}
\includegraphics[width=0.47\textwidth]{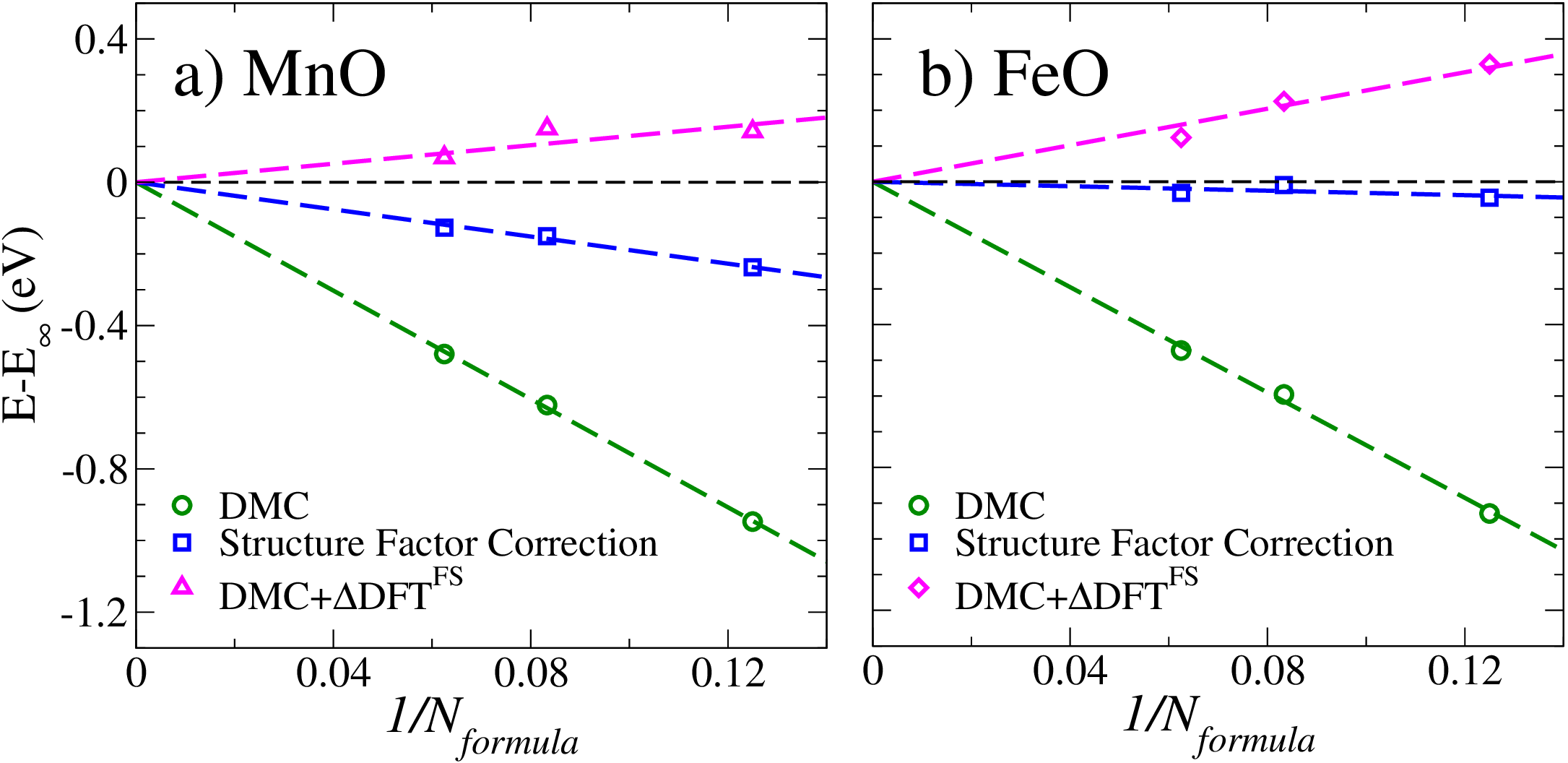}
\caption{(Color online)
 The relative FS errors (per number of formula units $Z$ in the supercell) for a) MnO in the B1 phase, and b) FeO in the iB8 phase at volumes of 21.7 ${\AA}^3$/MnO and 20.4 ${\AA}^3$/FeO, respectively. Results from the DMC calculations are indicated by green circles; corrected values using the structure factor method (taken from Ref. \onlinecite{mno,feo}) are shown by the blue squares.  DMC results corrected by  $\Delta\mathrm{DFT}^{\mathrm{FS}}$ 
 are indicated by red triangles.
 }
\label{fig:MnO-FeO}
\end{figure}

The last application is for crystalline MnO and FeO. These materials are challenging to simulate due to the complex interplay 
of strong correlation effects, p-d hybridization, and magnetism.
We apply the DFT FS correction directly to previous results from DMC calculations. \cite{mno,feo} Although the pseudopotentials used in our modified DFT calculations are different from those used in DMC, this has only a weak effect on the FS corrections. As seen in 
Fig.~\ref{fig:MnO-FeO}, the $\Delta\mathrm{DFT}^{\mathrm{FS}}$ correction 
greatly reduces the DMC FS error. 
In both materials, it somewhat over-corrects and approaches the infinite limit from above. The
structure factor correction of Chiesa \cite{chiesa,mno,feo}, also shown in the figure, somewhat under-corrects. For MnO the residual error is somewhat larger
than for $\Delta\mathrm{DFT}^{\mathrm{FS}}$, while it is smaller for FeO. 
A possible difficulty for $\Delta\mathrm{DFT}^{\mathrm{FS}}$ is the
 irregular supercells used for these materials. The $\Delta\mathrm{DFT}^{\mathrm{FS}}$ correction method is based on cubic supercells, parametrized
 by a single length scale $L$. This is a good approximation for other high symmetry systems such as fcc supercells. \cite{kzk} If the supercell shape is far from cubic and has strong anisotropy, however, the FS total energy will increase, resulting in a smaller FS correction compared to that of a cubic system. This is discussed further in the next Section.

 These applications demonstrate that the new FS error correction scheme can consistently remove most of the FS error and accelerate the size convergence of MB results to infinite size limit.

\section{Discussion}
\label{sec:discuss}

In this section we discuss several aspects of the new FS correction method, the 
parametrized LSDA functional, and possible further improvements.

In \mbox{Sec. \ref{sec:XC}}, the FS LSDA XC functional was separated into 
exchange $\varepsilon_x(r_s,L,\zeta)$  [Eqs.~(\ref{eq:EX-interpolation})-(\ref{eq:Exnew})] and 
correlation $\varepsilon_c(r_s,L,\zeta)$ [Eqs.~(\ref{eq:Ec}) and (\ref{eq:gr_Ec})]  contributions.
This facilitates the modification of existing DFT computer codes to generate the FS corrections. While useful,
this separation somewhat complicates the FS treatments, which are based on XC functionals derived from the HEG. 

For example, while the total FS HEG XC energy
scales as $1/L^3$, $\varepsilon_x(r_s,L,\zeta)$ scales as $1/L^2$, since the Hartree-Fock spectrum is gapless in this case.
The separation thus induces a  canceling $1/L^2$ term in $\varepsilon_c(r_s,L,\zeta)$.
Partly for this reason, we adopted the PZ spin interpolation formula given by Eq.~(\ref{eq:EX-interpolation}) for $\varepsilon_x(r_s,L,\zeta)$, rather than
the exact relation in Eq.~(\ref{eq:Ex}), since it makes the treatment of exchange consistent with that 
of the correlation, 
for which there is no analogy of  Eq.~(\ref{eq:Ex}), and ensures cancellation of the  $1/L^2$ dependence.
[As mentioned, in the limit $L \rightarrow \infty $, %is infinite,  
Eq.~(\ref{eq:EX-interpolation}) reduces to the exact relation in Eq.~(\ref{eq:Ex}).]
Since the FS exchange energy for spin unpolarized system is discontinuous, \cite{kzk}
another reason for using  Eq.~(\ref{eq:EX-interpolation}) is that it avoids multiple 
discontinuities in a partially polarized system, as discussed in Sec.~\ref{subsec:exchange}. 
The continuity density $\gamma_x$ in Eq.~(\ref{eq:Exnew}) was chosen for simplicity. 
It could be made polarization-dependent, but the expressions would be more complex and not as easy to implement into DFT codes. As discussed in Sec. \ref{subsec:xc-parameters}, FS corrections are not sensitive to the precise value of $\gamma_x$, especially for extended systems.

Similar considerations pertain
to the treatment of correlation.  Since $\varepsilon_c(r_s,L,\zeta)$ is continuous, it was easy to use different values for $\gamma_c(0)$ and $\gamma_c(1)$ in
Eq.~(\ref{eq:Ec}), as described in Sec.~\ref{subsec:xc-parameters}, when the PZ interpolation form is used.  
If the PW interpolation of Eq.~(\ref{eq:PW-interpolation}) were to be used, cancellation of the $1/L^2$ term
would require more complicated expressions.
Interpolation formulas, such as that of PZ, allow us to 
map spin unpolarized and fully-polarized results onto the correlation energy at an arbitrary polarization. For the fully polarized HEG, 
the available finite-size data from QMC were not as detailed as for the unpolarized case. Our fits of the fully-polarized $\varepsilon_c(r_s,L,p=1)$ correlation functional are based on AFQMC calculations of finite systems, mostly for $r_s$ greater than about 1.5 - 2.
While our correlation energy fits are acceptable (see Figs.~\ref{fig:Fit}  and \ref{fig:Ec-Inter}), more QMC calculations for small $r_s$ might lead to improvements.

Extensions of our FS error correction scheme are possible, for example based on orbital-dependent density functionals, \cite{Kumel2008} such as
hybrid DFT functional calculations. Hybrid functionals \cite{becke0,pbe0} augment the DFT exchange-correlation energy with an exact exchange term calculated from HF theory. Hybrid functionals have become 
one of the most promising approaches for overcoming some of the shortcomings of local or semilocal DFT approximations. \cite{Kumel2008}
Some complications would need to be addressed. Popular hybrid functionals such as PBE0 \cite{pbe0} use a fraction $\simeq 20$\,\% of exact exchange and DFT exchange for the remainder. In applications to semiconducting and insulating materials, the HF exchange converges as $1/L^3$ with PBC,  
since the single-particle spectrum has a gap. [The leading behavior is $1/L^3$ after a $1/L$ Madelung-like dependence is removed, \cite{gb,nguyen} as is done in QMC calculations. \cite{malliga}] Since DFT FS exchange scales as $1/L^2$ for the HEG, the fraction of the $1/L^2$
term in the FS correlation functional would have to be adjusted in order
to obtain a net $1/L^3$ behavior. The resulting  $\varepsilon_c(r_s,L,\zeta)$ would lead to a non-continuous 
behavior with increasing $r_s$, which might require further modifications of the fits. 

Finally, our FS XC functional is based on HEG calculations using cubic supercells of edge $L$. For non-cubic applications, $L$ is determined by the volume
of the supercell. In the HEG, spatially isotropic supercells are favored.  For example, a small cubic to  tetragonal distortion will cause the total energy to become more positive, scaling as $\simeq (1-C*\eta^2$), where
$C>0$ is a constant and $\eta$ describes the aspect ratio $c/a = 1+\eta$ of the tetragonal supercell.
Therefore our FS XC functional tends to provide an upper limit to the magnitude of the FS correction for non-cubic supercells. This may have contributed to the over-correction
seen in Fig.~\ref{fig:MnO-FeO}, where anisotropic MnO and FeO supercells were used.

\section{Summary}
\label{sec:summary}

Many-body calculational methods such as quantum Monte Carlo can potentially provide
accurate results for extended systems where electron interaction effects are important.
These calculations are computationally more costly than independent-electron calculations,
and are always accompanied by more significant FS errors. 
 In this work, we have provided a framework within DFT to understand and estimate 
 the FS errors 
 in MB calculations of spin-polarized systems. 
 A FS LSDA functional is parametrized which leads to a simple,
 post-processing correction method. The method is designed to approximately include 2B FS corrections in FS DFT calculations of systems with arbitrary spin polarization. The corrections to total energy, dissociation energy, and ionization energy calculations of several typical systems show a significant improvement on the convergence. 
The formalism of the method and the parametrization of the FS 
functional are presented in detail. 
The strengths and weaknesses of the approach are discussed.
There are several directions in which the method can be further developed. 
The method can be easily incorporated in any standard DFT computer program for 
periodic systems, and can be used to correct any MB results.
It is expected that the method will find applications in 
the study of a variety of realistic systems with magnetic ordering.

\acknowledgments

This work is supported by 
DOE (Grant no.~DE-FG02-09ER16046), NSF (Grant no.~DMR-1006217), and ONR (Grant no.~N00014-08-1-1235). Calculations were performed at the Center for Piezoelectric by Design. We thank 
S.~Sorella for helpful discussions, W. Purwanto for help with computing issues, and J. Koloren\v{c} and L. Mitas for sending us the computational details of Refs.~\onlinecite{mno,feo}.

\end{document}